# Connecting theory of plasmoid-modulated reconnection to observations of solar flares

Andrew Hillier[1,*]  and Shinsuke Takasao[2]

[1]Department of Mathematics and Statistics, University of Exeter, Exeter EX4 4QE, United Kingdom, and [2]Department of Earth and Space Science, Graduate School of Science, Osaka University, Toyonaka, Osaka 560-0043, Japan
*Corresponding author. Email: a.s.hillier@exeter.ac.uk

**Abstract**

The short timescale of the solar flare reconnection process has long proved to be a puzzle. Recent studies suggest the importance of the formation of plasmoids in the reconnecting current sheet, with quantifying the aspect ratio of the width to length of the current sheet in terms of a negative power $\alpha$ of the Lundquist number, that is, $S^{-\alpha}$, being key to understanding the onset of plasmoids formation. In this paper, we make the first application of theoretical scalings for this aspect ratio to observed flares to evaluate how plasmoid formation may connect with observations. For three different flares that show plasmoids we find a range of $\alpha$ values of $\alpha = 0.26$ to $0.31$. The values in this small range implies that plasmoids may be forming before the theoretically predicted critical aspect ratio ($\alpha = 1/3$) has been reached, potentially presenting a challenge for the theoretical models.

## Introduction

Solar flares, large releases of energy from the solar corona, are driven by a process called magnetic reconnection (e.g., Priest, 2014). This is where free energy stored in the magnetic field is released as thermal and kinetic energy through the magnetic field changing its connectivity (e.g., Yamada et al., 2010). Observations of flares show that the energy release takes place on a timescale of hours (e.g., Fletcher et al., 2011; Shibata & Magara, 2011). However, this timescale is much shorter than the timescale to diffuse the magnetic field in the solar corona of $10^6$ years (e.g., Shibata & Magara, 2011).

Understanding the short timescales, compared to the diffusion time, of solar flares has presented a theoretical challenge for many years. The first major step forward in explaining flare energy release was the Sweet–Parker reconnection model (Parker, 1957; Sweet, 1958), a steady-state model where flows bring magnetic field into a region of high current (called a current sheet), where it is annihilated, leading to heating and driving jets of material ejected from the reconnection region. In this model, the reconnection rate scales as $S^{-1/2}$, where $S$ is the Lundquist number defined as $S \equiv LV_A/\eta = \tau_\eta/\tau_A$ with $L$ the current sheet half-length, $V_A$ the Alfvén speed, $\eta$ the magnetic diffusivity, and $\tau_\eta$ and $\tau_A$ the diffusion and Alfvén times. With timescales for reconnection in this model scaling as the inverse root of the Lundquist number, with $S > 10^{12}$ in the solar corona, this model is still unable to explain the short timescales of solar flares.

To bridge this gap in timescales, it was proposed that the development of plasmoids in a reconnecting current sheet through the tearing instability (Furth et al., 1963) could play an important role in breaking up the current sheet and driving fast reconnection (e.g., Loureiro et al., 2007; Shibata & Tanuma, 2001). This would be either through creating a turbulent current sheet or through plasmoid dynamics locally thinning the current sheet to scales where kinetic effects can anomalously enhance the magnetic diffusivity (e.g., Zweibel & Yamada, 2009). Subsequently, observations have shown what appear to be plasmoids developing in a flare current sheet (e.g., Takasao et al., 2012) supporting this idea. Figure 1a shows the flare observed by Takasao et al. (2012) using the Atmospheric Imaging Assembly (AIA; Lemen et al., 2012) on the Solar Dynamics Observatory with a zoomed image of the current sheet showing the plasma blobs interpreted as plasmoids. Panel (b) of that figure gives a schematic diagram that explains how the observations connect to the formation of plasmoids in a reconnecting current sheet.

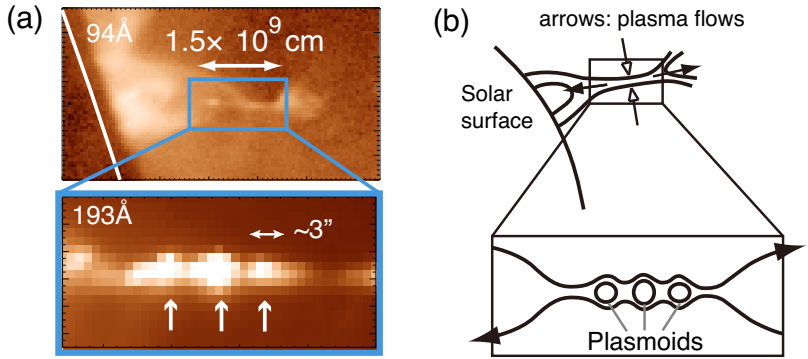

**Figure 1.** Image of a solar flare observed on August 18, 2010 with the Atmospheric Imaging Assembly (panel [a]). The zoomed image shows plasma blobs formed in the plasma sheet. Panel (b) presents a schematic diagram of these observations where the plasma sheet is understood as a current sheet with the plasma blobs interpreted as plasmoids.

To understand how the tearing instability develops in a reconnecting current sheet, it is standard practice to rescale the growth rates and wavenumbers to calculate them in terms of the half-length of the current sheet and not the half-width (which is used normally when calculating the growth rate of the instability; e.g., Furth et al., 1963). To do this, the aspect ratio $a/L$ (with $a$ the half-width of the current sheet and $L$ the half-length of the current sheet) is given to scale as $S^{-\alpha}$ (e.g., MacTaggart, 2020). After performing this rescaling, the instability is often known as the plasmoid instability. In a Sweet–Parker current sheet ($\alpha = 1/2$), the maximum growth rate of the instability scales as $S^{1/4}$ (e.g., Loureiro et al., 2007). However, as explained by Pucci and Velli (2014), the current sheet would become unstable to plasmoid formation before it has thinned/stretched to the Sweet–Parker aspect ratio. They proposed that $\alpha = 1/3$ is the correct scaling to expect as this is the aspect ratio where the tearing timescale becomes equal to the timescale for the ejection of a plasmoid (i.e., the Alfvén time).

The aspect ratio of $\sim S^{-1/3}$ has been found to be important for triggering the onset of plasmoid formation in a number of analytical and numerical studies (e.g., Comisso et al., 2016; Huang et al., 2017), but the question still remains as to what is happening in observed astrophysical systems. In this paper, we will investigate whether the magnetic reconnection behind observed solar flares can be understood through the plasmoid instability paradigm. We use some simple scaling laws to investigate the value of $\alpha$ required to explain the development of plasmoids that have been observed in solar flares, and connect these values with the current theoretical understanding.

### Scaling laws and their application to observations

For the tearing instability, there are well-established derivations of the most unstable mode of the instability (e.g., Tajima & Shibata, 2002). These give

$$\frac{a}{V_A}\sigma_{\max} = C_1 S^{*-1/2}, \qquad (1)$$

$$ak_{\max} = C_2 S^{*-1/4}, \qquad (2)$$

where $k_{\max}$ is the most unstable wavenumber of the system, $\sigma_{\max}$ is the corresponding growth rate, $S^* \equiv aV_A/\eta$ (the Lundquist number defined by the current sheet half-width $a$), and $C_1$ and $C_2$ are constants of order 1. For example, for a Harris current sheet, they become $C_1 \approx 0.62$ and $C_2 \approx 1.36$.

Following the arguments for the plasmoid instability (e.g., MacTaggart, 2020), we set that the half-width and half-length of the current sheet are connected by $a/L = S^{-\alpha}$. We can rescale the maximum growth rate and the most unstable wavelength to be in terms of $S$ and not $S^*$. For the growth rate, we have

$$\sigma_{\max} = \frac{V_A}{a} C_1 S^{*-1/2} = C_1 \frac{V_A}{L}\frac{L}{a} S^{-1/2}\left(\frac{a}{L}\right)^{-1/2} = C_1 \frac{V_A}{L} S^{(3\alpha-1)/2}. \qquad (3)$$

For the corresponding wave number, we have

$$k_{\max} = \frac{1}{a}C_2 S^{*-1/4} = C_2 \frac{1}{L}\frac{L}{a} S^{-1/4}\left(\frac{a}{L}\right)^{-1/4} = C_2 \frac{1}{L} S^{(5\alpha-1)/4}. \qquad (4)$$

Using Equation (4), and taking that $C_2 \approx 1$, we can rearrange to solve for $\alpha$, that is,

$$\alpha = \frac{4}{5}\left(\frac{\log_{10}(k_{\max}L)}{\log_{10}(S)} + \frac{1}{4}\right). \qquad (5)$$

It is these relations that we will apply to the flare observations to determine the value of $\alpha$.

The hypothesis we will test is whether, as expected from reconnection theory, solar flare observations present a consistent value of $\alpha$. If found, this would provide further evidence of the importance of the tearing instability in solar flare reconnection.

### Application to observed plasmoids

In this section, we analyze the plasmoids observed in three separate flares (displaying notably different scales). The data for the three flares we use are presented in Takasao et al. (2012), Milligan et al. (2010), and Patel et al. (2020), respectively. Below, we detail the key characteristics of these different observations, and then summarize the key quantities in Table 1, where the estimated $\alpha$ value is also presented.

The observations of Takasao et al. (2012) show the development of a long, thin current sheet, in which plasma blobs develop and are ejected. The outflow velocity (a good proxy for the Alfvén speed, e.g., Parker, 1957) was observed to be 220–460 km/s. Here, we take the largest value $4.6 \times 10^5$ m/s to be the representative value of the Alfvén speed. The half-length of the current sheet was observed to be at least $7 \times 10^6$ m. The typical size of the observed plasma blobs, interpreted to be plasmoids due to their movement along the current sheet, was $2.9 \times 10^6$ m (the interpretation as plasmoids was further supported by radio observations that detected the signatures of electron acceleration associated with their movement [Takasao et al., 2016]). This implies a wavenumber of $3 \times 10^{-6}$/m. Taking the temperature to be $10^6$ K, we expect the magnetic diffusivity to be 1 m$^2$/s, meaning that we have $S = 3.2 \times 10^{12}$.

**Table 1.** Key characteristics of the current sheet and plasmoids including the half-length of the current sheet, the estimated Alfvén speed, the characteristic plasmoid size, and the Lundquist number and $\alpha$ value calculated from these measurements for the different observed flares

| Obs. date | Current sheet half-length (m) | Alfvén speed (m/s) | Plasmoid size (m) | S | $\alpha$ |
|---|---|---|---|---|---|
| August 18, 2010[a] | $7 \times 10^6$ | $4.6 \times 10^5$ | $2.9 \times 10^6$ | $3.2 \times 10^{12}$ | 0.28 |
| January 25, 2007[b] | $4.4 \times 10^7$ to $1.1 \times 10^8$ | $4.6 \times 10^5$ to $1.3 \times 10^6$ | $2.5 \times 10^7$ | $2.2 \times 10^{13}$ to $1.4 \times 10^{14}$ | 0.26–0.28 |
| September 10, 2017[c] | $5.4 \times 10^7$ | $4.3 \times 10^5$ | $5.65 \times 10^6$ | $2.3 \times 10^{13}$ | 0.31 |

[a]Data extracted from Takasao et al. (2012).
[b]Data extracted from Milligan et al. (2010).
[c]Data extracted from Patel et al. (2020).

The observations of Milligan et al. (2010) show a plasmoid observed in hard X-ray by RHESSI (Reuven Ramaty High Energy Solar Spectroscopic Imager). From the observations presented in Figure 3 of Milligan et al. (2010), it seems reasonable to take a plasmoid (Source A in that figure) to be of size $\sim 35$ arcsec ($\sim 2.5 \times 10^7$ m), which becomes the wavelength used for our analysis. We can also make an estimate of the half-length of the current sheet. First, we can consider as a lower estimate the distance between the center of the two hard X-ray sources (again as shown in Figure 3 of Milligan et al., 2010), which is $\sim 60$ arcsec ($\sim 4.4 \times 10^7$ m). Alternatively, we can take as an upper limit of the length the distance between the flare arcade and the inner edge of the COR2 Coronagraph on the STEREO B satellite (Kaiser et al., 2008), which is $\sim 300$ arcsec, which gives an upper estimate of the half-length of 150 arcsec ($\sim 1.1 \times 10^8$ m). From these observations, it is not possible to directly estimate the Alfvén speed. We can give an upper estimate for the Alfvén speed from the CME speed, which is $1.3 \times 10^6$ m/s (the SOHO/LASCO CME Catalog; Yashiro et al., 2004), although we also look at the effect of using the slower speed of $4.6 \times 10^5$ m/s as found in the study of Takasao et al. (2012). Again, we take $\eta = 1$ m$^2$/s. These values give a range of Lundquist numbers between $S = 2.2 \times 10^{13}$ and $1.4 \times 10^{14}$.

The observations of Patel et al. (2020) show an evolving current sheet, which produces many plasmoids of varying size traveling at varying speeds. These plasmoids are observed in the lower solar corona by AIA and further out by COR2. To distill the observations into the key set of numbers we require, we focus on the observed plasmoids as seen by AIA. Taking the Alfvén speed as the approximate upper limit of the observed plasmoid velocity, we use a value of $4.3 \times 10^5$ m/s. The observed plasmoid width of $\sim 5.65 \times 10^6$ m is used as the wavelength of the tearing instability. The estimate for the half-length of the current sheet can be estimated from their Figure 11, where the stagnation height of the plasmoids can be determined. Subtracting the height of the flare arcade leads to a height of $\sim 75$ arcsec ($5.4 \times 10^7$ m). Again, we take $\eta = 1$ m$^2$/s.

The $\alpha$ values found for these observations are displayed in the rightmost column of Table 1, giving a range between 0.26 and 0.31. Although there is naturally some uncertainty in the number of the parameters used to calculate the $\alpha$ values, and this is potentially responsible for some of the spread observed, in general the widely different scales of the observations are presenting $\alpha$ values that are in a relatively small range. Considering the range of Lundquist numbers by one to two orders of magnitude, as well as the order of magnitude range in plasmoid size and current sheet lengths, this provides evidence that the theoretical understanding of how plasmoids are formed in flaring current sheets is consistent with the observations.

To highlight how robust the values calculated for $\alpha$ actually are, we can look at what happens if we take into account some of the uncertainties in our estimates for various quantities and apply these to the $\alpha$ value for the flare observed by Takasao et al. (2012). First, we can take $C_1$ to be 1.36. In this case, we find a value of $\alpha = 0.27$. Alternatively, we can assume that the half-length of our current sheet has been underestimated due to projection effects. Making $L$ to be 30% larger (and with it $S$ to be $1.3^2$ larger due to the effect of the projection effects on the estimate of the Alfvén speed), we find $\alpha = 0.28$. Finally, for the diffusion, we had assumed that the temperature of the medium was $10^6$ K, but $\eta \propto T^{-3/2}$ and the temperature in the current sheet is likely to be up to one order of magnitude hotter than the temperature assumed here. Taking $T = 10^7$ K, we find $\alpha = 0.27$. Further uncertainties still exist. The question of accurate determination of the current sheet length (where the observed length may also contain the reconnection jets as well as the current sheet) also leads to uncertainty. However, for the flare observed by Takasao et al. (2012), reducing the length by a factor of 2 results in a small reduction of $\alpha$ to $\alpha = 0.26$. Moreover, the observed plasmoid size may be overestimated as the observations are likely to show the later state of a plasmoid once it has accumulated more flux. Taking a plasmoid to initially be only half the observed size results in $\alpha = 0.29$. Therefore, we can conclude that reasonable levels of uncertainty do not result in large variation in the calculated value of $\alpha$.

Determining an $\alpha$ value has a particular consequence, and it implicitly makes a prediction for the thickness of the flare current sheet. Looking at the value obtained for the flare studied by Takasao et al. (2012), the calculated value of $\alpha$ implies that the actual thickness of the flare current sheet is $a \approx 2.2 \times 10^3$ m. This is approximately 2.5 orders of magnitude smaller than the observed half-thickness of the plasma sheet in the observations of $7 \times 10^5$ m (Takasao et al., 2012). However, this can be easily explained if the current sheet is not exactly aligned with the line of sight, meaning that its depth is projected to look like its width, or as a result of a thermal halo formed around the current sheet through heat conduction (Forbes & Malherbe, 1991; Takasao et al., 2015; Yokoyama & Shibata, 2001). In this case, the measured value of $\alpha$ would imply a timescale for the growth of the tearing instability of $\approx 150$ s. This is much shorter than the timeframe of the flare observations, meaning that for the $\alpha$ value we find it would be reasonable for plasmoids to develop during the course of the observations.

## Discussion

There have been many attempts in recent years to understand the connection between the fast, bursty reconnection observed in astrophysical systems through the plasmoid instability. These studies have made great progress through analytic theory and through numerical modeling. In this paper, we have extended these studies by looking at observational data of three solar flares and showing that it is consistent with an aspect ratio of $a/L = S^{-0.26}$ to $S^{-0.31}$.

The key finding for the solar flares we have studied is that the value for $\alpha$ ($\alpha = 0.26$ to $0.31$) is relatively close to the theoretical predictions of Pucci and Velli (2014) (e.g., $\alpha = 1/3$). However, it is important to note that this small difference is in fact somewhat difficult to reconcile through observational error due to the lack of sensitivity of $\alpha$ to reasonable estimates of the errors in the parameters used. There will always be some uncertainty with the value of $\alpha$ that cannot be quantified; for example, Huang et al. (2017) found in their numerical simulations that the wavenumber that ultimately grew was a factor of 3 to 6 smaller than the most unstable mode. If we take that the measured plasmoids are from a wavenumber six times smaller than the most unstable mode, we then find $\alpha = 0.31$ to $0.35$, that is, it gives the predicted scaling of $\alpha = 1/3$ by Pucci and Velli (2014), but it is not possible to prove with current observations that this process is happening in solar flare reconnection.

If we consider how a plasmoid may be formed when the timescale for its growth is longer than the expected timescale of the ejection, there may be some aspect of the reconnection flow that allows this to happen. For example, this may be occurring through plasmoid formation around the stagnation point of the flows into and out of the current sheet, allowing the first plasmoid to form in the current sheet where it takes significantly longer to eject allowing them to grow. This is an area for future investigation.

Magnetic reconnection is an important physical process in many astrophysical and space systems for driving the quick release of energy stored in magnetic fields. Therefore, quantifying how observed magnetic reconnection fits into current models of magnetic reconnection is an important topic of research. The methods laid out in this paper should be applicable for any observed reconnection region where the plasmoids have been observed. Finding further observations, in any system, to see if a similar value of $\alpha$ is consistently found would be an important future step.

**Acknowledgment.** A.H. would like to acknowledge the lectures of Dr. David MacTaggart at the Advanced Topics in MHD Summer School held in CISM, Udine, which led to the idea for this paper.

**Data availability statement.** The data used in this paper are available already in other published works.

**Author contributions.** A.H. designed the study and wrote the manuscript. S.T. provided feedback and guidance on the flare observations and application of the theory.

**Funding statement.** A.H. is supported by STFC Research Grant No. ST/V000659/1. S.T. is supported by JSPS KAKENHI Grant Nos. JP22H00134, JP22K14074, and JP21H04487.

**Conflict of interest.** Both authors have no conflicts of interest to declare.

# Peer Reviews

**Reviewing editor:** Prof. Stefano Camera, PhD

Universita degli Studi di Torino, Physics, Via Pietro Giuria, 1, Torino, Italy, 10124

Minor revisions requested.

doi:10.1017/exp.2022.23.pr1

## Review 1: Connecting Theory of Plasmoid-modulated Reconnection to Observations of Solar Flares

**Reviewer:** Dr. PengFei Chen

Date of review: 22 September 2022

**Conflict of interest statement.** Reviewer declares none.

## Comment

Comments to the Author: In this manuscript the authors tried to connect observational parameters of solar flares to the scaling law of magnetic reconnection, and it seems that the scaling law is quite similar to what proposed earlier via numerical simulations. If the topic fits into the theme of the journal, I'd like to recommend it to be published after revisions.

Major issues:

1. Personally I tend to think that the 1.5*10^9 cm in Fig. 1 is not the length of the current sheet. It might consists of a much shorter current sheet in the middle and long bidirectional outflows both upward and downward. In this case L might be much smaller. Its effect on alpha should also be discussed on p.4.

2. In this paper, the magnetic diffusivity is taken to be the classical value. However, it has been proposed that anomalous resistivity is required for MR to occur, and particle simulations indeed showed that the diffusivity is enhanced by ~6 orders of magnitude during reconnection. In this case, S would be substantially smaller, and alpha would be much larger according to Eq. (5).

Minor issues:

1. In several places, "magnetic diffusion" should be "magnetic diffusivity";
2. In several places, "Arcsec" should be "arcsec";
3. P. 1: "is unable to still" --> "is still unable to";
4. Figure1 panel (a) --> Figure 1(a);
5. table 1 -->Table 1;
6. The X8.2-class solar flare on 2017-09-10 can be used as the 4th example if possible.

## Score Card

### Presentation

**3.7 /5**

| | |
|---|---|
| Is the article written in clear and proper English? (30%) | 4/5 |
| Is the data presented in the most useful manner? (40%) | 4/5 |
| Does the paper cite relevant and related articles appropriately? (30%) | 3/5 |

### Context

3.8 /5

| | |
|---|---|
| Does the title suitably represent the article? (25%) | 5/5 |
| Does the abstract correctly embody the content of the article? (25%) | 4/5 |
| Does the introduction give appropriate context? (25%) | 4/5 |
| Is the objective of the experiment clearly defined? (25%) | 2/5 |

### Analysis

3.8 /5

| | |
|---|---|
| Does the discussion adequately interpret the results presented? (40%) | 4/5 |
| Is the conclusion consistent with the results and discussion? (40%) | 4/5 |
| Are the limitations of the experiment as well as the contributions of the experiment clearly outlined? (20%) | 3/5 |

# Review 2: Connecting Theory of Plasmoid-modulated Reconnection to Observations of Solar Flares

**Reviewer:** Dr. Yi-Min Huang

Princeton University, United States

Date of review: 26 August 2022

**Conflict of interest statement.** Reviewer declares none.

## Comment

Comments to the Author: The authors' approach of using the observed sizes of plasmoids and the rescaled fastest-growing wavenumber to infer the aspect ratio of the current sheet is quite ingenious. The authors have also discussed several possible uncertainties in their method. In addition to those uncertainties, I would like to point out another cause of errors that the authors have not addressed. Simulations have shown that small plasmoids tend to coalesce and form larger plasmoids. Moreover, plasmoid size grows over time as they chew magnetic flux through reconnection. Plasmoids could be smaller than the observed ones when they were first born but only become visible when they grow larger through coalescence or reconnection. Therefore, the observed plasmoid sizes may be viewed as an upper bound for the plasmoid instability wavelength. Future solar imagers with higher resolution may help to resolve this problem.

Minor issues:

(a) Page 2, last paragraph: The reference Comisso et al. (2016) is a purely analytic paper and contains no simulations. It is true that $S^{-1/3}$ appears in the aspect ratio derived by Comisso et al., but there is an additional logarithmic factor that depends on S and the initial perturbation amplitude.

(b) Second paragraph of Discussion, line 7: "siz" appears to be a typo.

## Score Card

### Presentation

5.0 /5

| | |
|---|---|
| Is the article written in clear and proper English? (30%) | 5/5 |
| Is the data presented in the most useful manner? (40%) | 5/5 |
| Does the paper cite relevant and related articles appropriately? (30%) | 5/5 |

### Context

5.0 /5

| | |
|---|---|
| Does the title suitably represent the article? (25%) | 5/5 |
| Does the abstract correctly embody the content of the article? (25%) | 5/5 |
| Does the introduction give appropriate context? (25%) | 5/5 |
| Is the objective of the experiment clearly defined? (25%) | 5/5 |

## Analysis

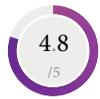

4.8 /5

| | |
|---|---|
| Does the discussion adequately interpret the results presented? (40%) | 5/5 |
| Is the conclusion consistent with the results and discussion? (40%) | 5/5 |
| Are the limitations of the experiment as well as the contributions of the experiment clearly outlined? (20%) | 4/5 |